\documentclass[twocolumn,trackchanges]{aastex7}

\newcommand{\Msun}{M$_\odot$} 

\newcommand{\mesa}{\textsc{Mesa}}

\usepackage[T1]{fontenc}
\usepackage{amsmath}

\begin{document}

\title{MESA-QUEST: Modeling Quasi-Stars in MESA}

\author[orcid=0009-0002-5773-3531,gname=Claire,sname=Campbell]{Claire B. Campbell}
\affiliation{Illinois State University, Department of Physics}
\email[show]{cbcamp1@ilstu.edu}

\author[orcid=0009-0007-4394-3366, gname='Andy', sname='Santarelli']{Andrew D. Santarelli}
\affiliation{Illinois State University, Department of Physics}
\affiliation{Yale University, Department of Astronomy}
\email[show]{andy.santarelli@yale.edu}  

\author[orcid=0009-0006-2122-5606,gname='Matt', sname='Caplan']{Matthew E. Caplan} 
\affiliation{Illinois State University, Department of Physics}
\email[show]{mecapl1@ilstu.edu}

\begin{abstract}

Supermassive black hole formation remains an unsolved problem. Quasi-stars have been suggested as a viable heavy-seeding mechanism. In this work, we implement methods for modeling quasi-stars previously used with the Cambridge \textsc{STARS} code into the 1D stellar evolution code \mesa. The computational capabilities of \mesa\ allow for more detailed simulations of quasi-star evolution due to its modularity and the ease of implementing of new physical processes and controls.  Our implementation, the \mesa\ Quasi-star Evolutionary Simulation Toolkit (MESA-QUEST),  is available in a publicly accessible repository.
\end{abstract}

\keywords{\uat{supermassive black holes}{1663} --- \uat{black holes}{162} --- \uat{stellar evolutionary models}{2046}}

\section{Introduction} \label{sec:intro}

Recent JWST observations of early-universe supermassive black holes (SMBHs), such as the $\sim4 \times 10^7$\ \Msun\ quasar in UHZ1 \citep{Natarajan_2024}, motivate work studying their early-time formation pathways. Direct collapse, which can occur when a massive cloud of low-metallicity pre-galactic gas monolithically collapses to form a central black hole, may be a viable means of seeding SMBHs \citep{Begelman_2006}. This may be observable as a `quasi-star', a hypothetical star-like object where an extended spherical envelope of gas is supported by the central black hole's accretion luminosity rather than nuclear fusion \citep[\emph{e.g.}][]{Begelman_2008}. 

Previous works have computationally modeled quasi-stars to determine their observable properties and evaluate their potential as a mechanism to seed SMBHs \citep{Ball_2011, Ball_2012, Coughlin_2024}. In this work, we develop new capabilities for simulating quasi-star models using the 1D stellar evolution code \mesa\ \citep{Paxton_2011}. We are motivated to extend \mesa\ because it is an open-source, up-to-date stellar evolution code with a wide variety of pre-existing capabilities. \mesa's capabilities enable us to more easily include physical processes absent from past simulations such as envelope accretion, mass-loss from winds, and photon-trapping.
This requires a number of changes to the code itself and adjustments of the input parameters within \mesa, which we discuss in the following sections. We validate our \mesa\ implementation by reproducing the fiducial models from \cite{Ball_2011}.

Our code and examples are available in a public repository.\footnote{ \href{http://www.github.com/andysantarelli/MESA-QUEST}{http://www.github.com/andysantarelli/MESA-QUEST}} All new physics are implemented in the \textit{run\_star\_extras} file. The provided \textit{inlist\_project} will simulate the model shown in Fig. \ref{fig:comp}. The \textit{x\_ctrl} parameters within this file control dimensionless efficiency parameters.

\section{Implementation} \label{sec:models}

\begin{figure*}[hbt!]
    \centering
    \includegraphics[clip, width=\linewidth]{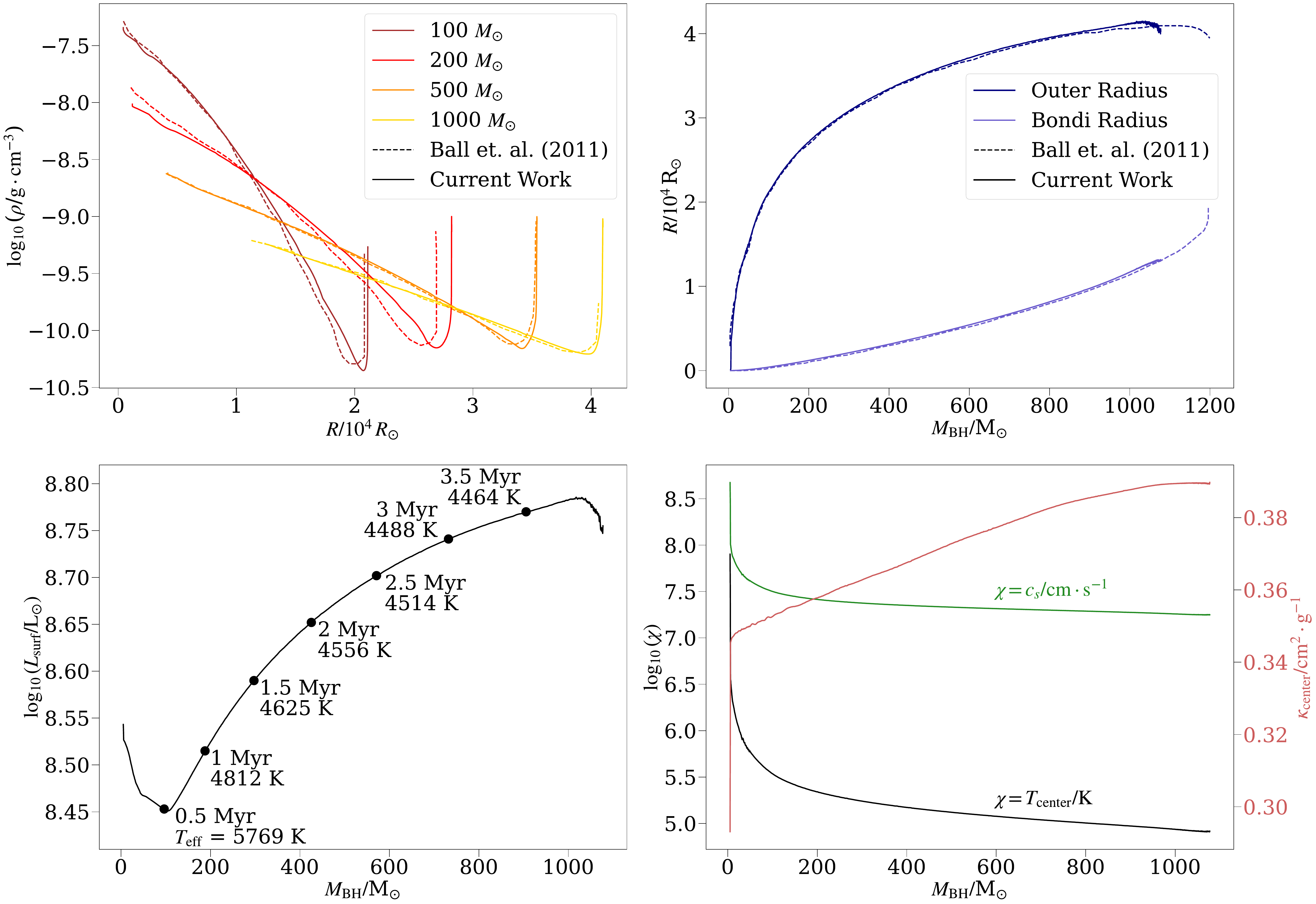}
    \caption{\textbf{ \textbf{\textsc{Mesa}} and Ball (2011) Comparison:} (Top) Density as a function of radial profile (left) and outer and Bondi radius as a function of black hole mass (right) from \cite{Ball_2011} (dashed) and our fiducial \mesa\ model (solid) using $(\epsilon,\,\eta)=(0.1,\,0.1)$. For the density, these profiles are chosen to match available data from \cite{Ball_2011} and are not evenly spaced in time. (Bottom) Surface (left) and core (right) properties of our quasi-star as functions of black hole mass, including model ages. } 
    \label{fig:comp}
\end{figure*}

Our implementation follows from the methods introduced in \cite{Ball_2011} and \cite{Ball_2012}. We treat the central black hole as a point mass at the origin. To prevent singularities, we adjust the center-most point of the simulation to $r_0$, beyond the Schwarzschild radius of the black hole. The mass contained within the sphere of radius $r_0$ is $M_0$, including the black hole and some infalling mass. These changes are given by

\begin{align}
    r_0 &= r_B = \frac{2GM_\textrm{BH}}{c_s^2}, \\
    M_0 &= M_\textrm{BH} + M_\textrm{cav},
\end{align}

\noindent where $r_B$ is the central black hole's Bondi radius, $M_{\rm BH}$ is the mass of the black hole, $c_s$ is the core speed of sound, and $M_\textrm{cav}$ is the mass contained within the inner boundary radius and comes from the estimation of the density profile where $\rho \propto r^{-3/2}$. With these adjustments, \mesa\ now integrates the equations of stellar structure outward from $r_0$ and $M_0$ rather than 0. 

The accretion rate and the corresponding convectively-limited luminosity are given by

\begin{align}
    \dot M_\textrm{BH} &= 16\pi \frac{\eta}{\epsilon \Gamma} \frac{(GM_\textrm{BH})^2}{c_s c^2} \rho \\
    L_\textrm{BH}  &= \frac{\epsilon}{1-\epsilon} \dot M_\textrm{BH} c^2 
\end{align}

\noindent where $\eta$ and $\epsilon$ are the convective and radiative efficiencies respectively, both taken to be 0.1, and $\Gamma$ is the adiabatic index which is evaluated self-consistently within \mesa. This results in a luminosity close to the Eddington limit for the entire star such that 

\begin{equation}
    L_{\star, \rm{Edd}} = 4\pi \frac{c}{\kappa}GM_{\star}.
\end{equation}

Extreme conditions surrounding the BH result in data noise close to $r_0$, particularly in $c_s$, when solving equations of stellar structure. To help stabilize this noise, we average the sound speed across the innermost 50 zones ($<1\%$ of the object by mass). Additionally, we put an upper limit of $5\%$ on the fractional change in $r_B$ at each timestep as a safety net to prevent large, non-physical changes caused by the noise. For our fiducial models below we substitute the default mixing length theory in \mesa\ for that of \cite{Bohm-Vitense_1958}, denoted `ML1', and use a mixing length parameter $\alpha_{MLT}=2$ to match the methods of \cite{Ball_2011}.  

Due to their high masses and core densities, quasi-stars are prone to the general relativistic instability and thus require corrections to account for these effects \citep{Herrington:2022dbu}. This is handled in \mesa\ by applying the Tolman-Oppenheimer-Volkoff correction to the equation of hydrostatic equilibrium. We implement this through the \textit{run\_star\_extras} file, where \mesa\ applies the correction to each zone at every timestep. Additionally, we implement the Ledoux criterion for convective stability, using the default semi-convective mixing efficiency parameter \citep{Herrington:2022dbu}.

To verify \mesa's capability to simulate quasi-stars, we compare our model to the fiducial model in \cite{Ball_2011}, modeled in the Cambridge \textsc{STARS} code \citep{STARS}. In Fig. \ref{fig:comp} we show the density profiles of our model as the central black hole grows (top left) and the position of the inner radius $r_0$ and the outer radius of the star (top right). Our model reaches a final BH mass within 10\% of the \cite{Ball_2011} model and follows a similar structural evolution. The surface luminosity (bottom left) grows steadily with age while the surface temperature gradually declines. The core conditions (bottom right) are numerically stable. 

\section{Conclusions} \label{sec:concl}

We have developed new capabilities for \mesa\ to simulate quasi-stars and demonstrated consistency with the literature. This will allow us to easily include previously omitted processes in our future work, such as mass loss from winds and mass gain from envelope accretion, to better determine the circumstances in which quasi-stars form heavy seeds. As \mesa\ is open source, it is straightforward to adapt our code to run with new accretion schemes that have been developed in recent years, which will be studied in future work.

\begin{acknowledgments}

Financial support for this publication comes from Cottrell Scholar Award \#CS-CSA-2023-139 sponsored by Research Corporation for Science Advancement. This work was supported by a grant from the Simons Foundation (MP-SCMPS-00001470) to MC. This research was supported in part by the National Science Foundation under Grant No. NSF PHY-1748958.

\end{acknowledgments}


\begin{thebibliography}{}
\expandafter\ifx\csname natexlab\endcsname\relax\def\natexlab#1{#1}\fi
\providecommand{\url}[1]{\href{#1}{#1}}
\providecommand{\dodoi}[1]{doi:~\href{http://doi.org/#1}{\nolinkurl{#1}}}
\providecommand{\doeprint}[1]{\href{http://ascl.net/#1}{\nolinkurl{http://ascl.net/#1}}}
\providecommand{\doarXiv}[1]{\href{https://arxiv.org/abs/#1}{\nolinkurl{https://arxiv.org/abs/#1}}}

\bibitem[{W.~H. Ball(2012)Ball}]{Ball_2012}
Ball, W.~H. 2012, PhD thesis, Apollo - University of Cambridge Repository, \dodoi{10.17863/CAM.16000}

\bibitem[{W.~H. Ball {et~al.}(2011)Ball, Tout, Żytkow, \& Eldridge}]{Ball_2011}
Ball, W.~H., Tout, C.~A., Żytkow, A.~N., \& Eldridge, J.~J. 2011, \bibinfo{title}{The structure and evolution of quasi-stars,} Monthly Notices of the Royal Astronomical Society, 414, 2751, \dodoi{10.1111/j.1365-2966.2011.18591.x}

\bibitem[{M.~C. Begelman {et~al.}(2008)Begelman, Rossi, \& Armitage}]{Begelman_2008}
Begelman, M.~C., Rossi, E.~M., \& Armitage, P.~J. 2008, \bibinfo{title}{Quasi-stars: accreting black holes inside massive envelopes,} Monthly Notices of the Royal Astronomical Society, 387, 1649, \dodoi{10.1111/j.1365-2966.2008.13344.x}

\bibitem[{M.~C. Begelman {et~al.}(2006)Begelman, Volonteri, \& Rees}]{Begelman_2006}
Begelman, M.~C., Volonteri, M., \& Rees, M.~J. 2006, \bibinfo{title}{Formation of supermassive black holes by direct collapse in pre-galactic haloes,} Monthly Notices of the Royal Astronomical Society, 370, 289, \dodoi{10.1111/j.1365-2966.2006.10467.x}

\bibitem[{E. {B{\"o}hm-Vitense}(1958){B{\"o}hm-Vitense}}]{Bohm-Vitense_1958}
{B{\"o}hm-Vitense}, E. 1958, \bibinfo{title}{{{\"U}ber die Wasserstoffkonvektionszone in Sternen verschiedener Effektivtemperaturen und Leuchtkr{\"a}fte. Mit 5 Textabbildungen},} \zap, 46, 108

\bibitem[{E.~R. Coughlin \& M.~C. Begelman(2024)Coughlin \& Begelman}]{Coughlin_2024}
Coughlin, E.~R., \& Begelman, M.~C. 2024, \bibinfo{title}{Quasi-stars as a Means of Rapid Black Hole Growth in the Early Universe,} The Astrophysical Journal, 970, 158, \dodoi{10.3847/1538-4357/ad5723}

\bibitem[{P.~P. Eggleton(1972)Eggleton}]{STARS}
Eggleton, P.~P. 1972, \bibinfo{title}{{Composition Changes During Stellar Evolution},} Monthly Notices of the Royal Astronomical Society, 156, 361, \dodoi{10.1093/mnras/156.3.361}

\bibitem[{N.~P. Herrington {et~al.}(2023)Herrington, Whalen, \& Woods}]{Herrington:2022dbu}
Herrington, N.~P., Whalen, D.~J., \& Woods, T.~E. 2023, \bibinfo{title}{{Modeling Supermassive Primordial Stars with MESA},} Mon. Not. Roy. Astron. Soc., 521, 463, \dodoi{10.1093/mnras/stad572}

\bibitem[{P. Natarajan {et~al.}(2023)Natarajan, Pacucci, Ricarte, Bogdán, Goulding, \& Cappelluti}]{Natarajan_2024}
Natarajan, P., Pacucci, F., Ricarte, A., {et~al.} 2023, \bibinfo{title}{First Detection of an Overmassive Black Hole Galaxy UHZ1: Evidence for Heavy Black Hole Seed Formation from Direct Collapse,} The Astrophysical Journal Letters, 960, L1, \dodoi{10.3847/2041-8213/ad0e76}

\bibitem[{B. {Paxton} {et~al.}(2011){Paxton}, {Bildsten}, {Dotter}, {Herwig}, {Lesaffre}, \& {Timmes}}]{Paxton_2011}
{Paxton}, B., {Bildsten}, L., {Dotter}, A., {et~al.} 2011, \bibinfo{title}{{Modules for Experiments in Stellar Astrophysics (MESA)},} \apjs, 192, 3, \dodoi{10.1088/0067-0049/192/1/3}

\end{thebibliography}

\end{document}